\documentstyle[preprint,eqsecnum,aps]{revtex}
\begin{document}
\draft
\preprint{HEP/123-qed}

\title{Angle-resolved photoemission study of $MX$-chain compound
[Ni(chxn)$_2$Br]Br$_2$: absence of spin-charge separation signals}

\author{Shin-ichi Fujimori, Akihiro Ino, and Testuo Okane}
\address{Synchrotron Radiation Research Center,
Japan Atomic Energy Research Institute,
SPring-8, Mikazuki, Hyogo 679-5148, Japan}

\author{Atsushi Fujimori}
\address{Synchrotron Radiation Research Center,
Japan Atomic Energy Research Institute,
SPring-8, Mikazuki, Hyogo 679-5148, Japan
\\and\\
Department of Complexity Science, University of Tokyo,
Bunkyo-ku, Tokyo 113-0033, Japan}

\author{Kozo Okada}
\address{Department of Physics, Faculty of Science, Okayama University,
Tsushima-naka, Okayama 700-8530, Japan}

\author{Toshio Manabe and Masahiro Yamashita}
\address{Department of Chemistry, Graduate School of Science,
Tokyo Metropolitan University, Hachioji, Tokyo 192-0397, Japan}

\author{Hideo Kishida and Hiroshi Okamoto}
\address{Department of Advanced Materials Science,
Graduate School of Frontier Science, University of Tokyo,
Bunkyo-ku, Tokyo 113-8656, Japan}

\date{\today}
\maketitle
\begin{abstract}
We report on the results of angle-resolved photoemission experiments on a
quasi-one-dimensional $MX$-chain compound [Ni(chxn)$_2$Br]Br$_2$
(chxn = 1$R$,2$R$-cyclohexanediamine), a one-dimensional Heisenberg system
with $S=1/2$ and $J \sim 3600$~K, which shows a gigantic non-linear optical effect.
A "band" having about 500~meV energy dispersion
is found in the first half of the Brillouin zone $(0\le kb/\pi <1/2)$,
but disappears at $kb / \pi \sim 1/2$.
Two dispersive features, expected from the spin-charge separation,
as have been observed in other quasi-one-dimensional systems like Sr$_2$CuO$_3$,
are not detected.
These characteristic features are well reproduced
by the $d$-$p$ chain model calculations with a small charge-transfer energy $\Delta$
compared with that of one-dimensional Cu-O based compounds.
We propose that this smaller $\Delta$ is the origin of the absence of clear spin- and
charge-separation in the photoemission spectra and strong non-linear optical effect
in [Ni(chxn)$_2$Br]Br$_2$.
\\
\end{abstract}

\narrowtext

Recently, one-dimensional (1D) strongly-correlated (SC) electron systems
have attracted much attention.
One of the most striking findings in these systems is the so-called spin-charge separation.
The spin and charge degrees of freedom are decoupled, and the low energy properties
of the material are governed by a spinless excitation called a holon
and a chargeless excitation called a spinon.
These elemental excitations were theoretically first predicted
by Lieb and Wu\cite{LiebWu}, and were experimentally observed by
recent angle-resolved photoemission spectroscopy (ARPES) experiments on SrCuO$_2$\cite{Kim},
Sr$_2$CuO$_3$\cite{Fujisawa}, and NaV$_2$O$_5$\cite{Kobayashi}. 
In addition to these purely academic interests, it is recently recognized that
these 1D-SC electron systems show strong non-linear optical effects.
Kishida {\it et al} \cite{Kishida} have observed a significantly enhanced third-order non-linear
dielectric susceptibility $\chi^{(3)}$ in Sr$_2$CuO$_3$ and
the halogen($X$)-bridged transition-metal($M$) compounds, designated as
the $MX$-chain compounds.
The large $\chi^{(3)}$ is indispensable for
opto-electronic switching, modulating and computing devices,
and its magnitude dominates their performances.
These 1D-SC electron systems may therefore become key materials
for future optical technology.

Among those 1D-SC electron systems, the $MX$-chain compounds with $M$=Ni, $X$=Br,
[Ni(chxn)$_2$Br]Br$_2$, (chxn = 1$R$, 2$R$-cyclohexanediamine)
have the largest value of $\max |{\rm Im} \chi^{(3)}|$,
and are good target materials to study the underlying physics of these properties.
In spite of these interesting properties of [Ni(chxn)$_2$Br]Br$_2$,
there have been few studies on its electronic structure.
A previous X-ray photoemission spectroscopy (XPS) study has revealed that
this material is a charge-transfer (CT) type insulator\cite{Okamoto2}.
However, the core-level experiment alone was insufficient to obtain
information about its 1D electronic structure
since the charge-transfer satellites in the core-level spectra are originated from
not only $p$-$d$ hybridization along the chain, but also that perpendicular to the chain.
In this paper, we have performed ARPES experiments
on this compound to reveal its electronic structure.
The result is compared with that of Cu-O based systems, such as Sr$_2$CuO$_3$,
and analyzed with calculations based on the $d$-$p$ chain model.

The basic crystal structure of [Ni(chxn)$_2$Br]Br$_2$ consists of
Ni(chxn)$_2$-Br chains and counter Br ions.
The chains are well separated by the counter Br atoms and (chxn) molecules,
and interaction between them is considered to be negligibly weak.
The N atoms of ligand (chxn) molecules are very close to Ni atoms,
leading to a strong crystal field on the Ni 3$d$ states.
The resulting electronic configuration of the Ni$^{3+}$ ion
is the filled $t_{2g}$ orbitals plus a singly occupied $e_g$
(3$d_{3z^2-r^2}$) orbital, $(t_{2g})^6(e_g)^1$.
The strong overlap between the Ni~3$d_{3z^2-r^2}$ orbital and
the halogen $p_{\rm z}$ orbital lead to a strong magnetic coupling along the chain and
the magnetic susceptibility $\chi$ is described by the Bonner-Fisher formula
with $S=1/2$ and $J \sim 3600$~K (=0.31~eV)\cite{Okamoto1}.
Thus this material can be viewed as a 1D Heisenberg chain system.

Single crystals of [Ni(chxn)$_2$Br]Br$_2$ were prepared by an electrochemical method.
ARPES experiments were performed using a spectrometer
equipped with a GAMMADATA-SCIENTA SES2002 electron analyzer and
a monochromatized GAMMADATA-SCIENTA vacuum ultraviolet (VUV) lamp.
Clean sample surfaces were obtained by {\it in-situ} cleaving
the sample parallel to the $bc$ plane.
In order to avoid charging effects, the sample was kept at room temperature.
The overall energy resolution was set at about 50~meV.
Binding energies are referred to the Fermi edge of an evaporated gold film.

During the course of the measurement, gradual changes in the line shape of
the photoemission spectra were observed.
These were not due to charging effects of the sample surface
since the line shape changes were not associated with rigid shifts of the spectra.
Further observations revealed that these changes accumulated with time
and were correlated with the strength of the VUV irradiation on the sample surface.
Therefore, we concluded that the changes were due to the degradation of
the sample surface caused by the irradiation, which had also been observed 
in the previous XPS measurements\cite{Okamoto2}.
We then dscided to use He II$\alpha$ ($h\nu=40.8$~eV) radiation for the present measurements,
whose intensities are about twenty times weaker than that of the often-used
He I$\alpha$ ($h \nu = 21.2$~eV) radiation.
In addition, we took all the spectra within twenty minutes.
We have carefully checked the time dependence of the spectral line shape,
and confirmed that no remarkable changes were observed within this period.
Another advantage of using He II radiation for the ARPES measurements
is that it is possible to observe the most of the first Brillouin zone of this material
without rotating the sample
because with the present analyzer and the He II$\alpha$ radiation, one can
detect the photoelectrons emitted about $\pm6$ degrees at the same time\cite{ARPES}.
This made it possible to measure all the spectra before the sample surface was damaged
by the VUV radiation.

Figure~\ref{ARPES} shows the ARPES spectra of [Ni(chxn)$_2$Br]Br$_2$.
The momentum scan was performed along the $b$-axis (chain axis).
All spectra have been normalized to the integrated intensities between $E_{\rm F}$ and 1.0~eV.
There is a weak but distinctive peak structure around 0.3-0.8~eV.
The peak is located at about 0.8~eV at $kb/\pi =0$, and moves toward
lower binding energy as $kb/\pi$ increases.
It reaches to about 0.3~eV at $kb/\pi=1/2$, and suddenly disappears with
additional increase in $kb$.
To see these features more clearly, we have derived the "image" of the band dispersion
from these spectra by taking second derivatives of smoothed ARPES spectra.
This method is very sensitive to peak structures.
Figure~\ref{image} shows their gray plot as a function of binding energy and momentum,
where the dark part corresponds to peak position.
We have tested different smoothing procedures in making images carefully,
and confirmed that the essential spectral feature does not depend on the procedures.
The above-mentioned spectral feature is more clearly understood from this image.
The sine-curve like dispersion is observed in the first half
($0\le kb/\pi < 1/2$) of the Brillouin zone,
but it suddenly disappears at $kb/\pi \sim 0.5$, and then
is not seen in the second half ($1/2\le kb/\pi < 1$) of the Brillouin zone.
The dotted white line is a guide to the eye.
The size of the energy dispersion is estimated to be about 500~meV.
In this scan, the spectra were measured only up to $kb/\pi\sim0.92$,
but spectra in the remaining part of the Brillouin zone were also been measured
in a different scan, and it is confirmed that there is no prominent feature
at $kb > 0.92$.
This result is qualitatively different from those of Cu-O based 1D-SC compounds
like SrCuO$_2$ or Sr$_2$CuO$_3$.
In the first half of the Brillouin zone, two bands originating from the spinon and holon
excitations are observed in the Cu-O based 1D-SC compounds,
while only one band is observed in [Ni(chxn)$_2$Br]Br$_2$.
Moreover, in the second half of the Brillouin zone, one band,
originating from a holon, is observed in the Cu-O based 1D-SC compounds,
while no band is observed in [Ni(chxn)$_2$Br]Br$_2$. 

In order to better understand the ARPES spectra of [Ni(chxn)$_2$Br]Br$_2$,
we have calculated the spectral function of these materials within the framework of
the $d$-$p$ chain model.
We have employed this model, because this model incorporates
large enough degrees of freedom to properly treat the CT-type system.
The more frequently used $t$-$J$ model can be derived from this model
in the limit of $U\to \infty$.
Moreover, the $d$-$p$ model can include photoemission cross section effects,
which cannot be taken into account in the $t$-$J$ model.
In this model, the $d$ and $p$ orbitals are arranged alternatively in a chain,
and hybridization between the nearest neighbor $d$ and $p$ sites is considered.
The model is expressed by the Hamiltonian
\begin{equation}
H_{dp} = \Delta \sum_{i \sigma}n_{i \sigma}^p + U \sum_i n_{i \uparrow}^d n_{i \downarrow}^d
- t \sum_{<ij>} (d_{i \sigma}^\dagger p_{j \sigma} + d_{i \sigma} p_{j \sigma}^\dagger ) ,
\end{equation}
where $n_{i \sigma}^d$ and $n_{i \sigma}^p$ are the number operators of the $d$ and $p$ holes
with spin $\sigma$ at site $i$, and $d_{i \sigma}$ and $p_{i \sigma}$ are
the annihilation operators of the $d$ and $p$ holes with spin $\sigma$ at a site $i$, respectively.
The sum $<ij>$ takes all combinations of the nearest-neighbor $d$ and $p$ sites.
$\Delta$ is the charge-transfer energy between $d$- and $p$-sites,
$U$ is the Coulomb interaction between $d$ electrons on the site,
and $t$ is the transfer integral between the nearest $d$ and $p$ states.
We have performed numerical diagonalization of the Hamiltonian matrix for a finite size cluster,
and calculated the single particle spectral function $A(k, \omega)$ expressed as 
\begin{equation}
A(k,\omega) = \sum_{\nu \sigma} |\langle \nu | a_{k \sigma}^{\dagger} | 0 \rangle|^2
{\rm \delta} ( \omega - E_\nu + E_0 ) ,
\end{equation}
where $a_{k \sigma}^{\dagger} = L^{-1/2} \sum_i d_{i \sigma}^{\dagger}e^{ikR_i}$ or
$L^{-1/2} \sum_i p_{i \sigma}^{\dagger}e^{ikR_i}$ is a creation operator
for a $d$ or $p$ hole, respectively.
$| \nu \rangle$ is the $\nu$-th eigenvector for the photoemission final state
with eigenvalue $E_\nu$.
$| 0 \rangle$ denotes the ground state eigenvector with eigenvalue $E_0$.
This method is exact, but requires the diagonalization of a huge Hamiltonian matrix.
Here, we took an 8-molecule $d$-$p$ chain model with four holes per spin.
This cluster size is small, but five independent $k$-points
in the Brillouin zone can be obtained, and therefore it is possible
to simulate the dispersive features of the experimental spectra.
In these calculations, the $p$ hole levels are empty and
the $d$ hole levels are half-filled in the initial state.
To check the validity of these parameters, the exchange interaction $J$ between the $d$ sites,
which can be compared with the experimental value, was also evaluated in the following ways.
We considered a $2\frac{1}{2}$-molecule $d$-$p$ cluster ($p$-$d$-$p$-$d$-$p$),
and calculated the total energy for the total spin $S_z=0$ and $S_z=1$ with these parameters.
The exchange energy is then evaluated by $J=E(S_z=1)-E(S_z=0)$.

First we present the calculations for a Cu-O chain to reproduce the previous theoretical
and experimental results\cite{Kim,Fujisawa}.
Figure~\ref{calc}(a) and (b) show the results of the calculations
for $t=1$, $\Delta=2.4$, and $U=7$~eV.
These parameters are similar to those used in the calculation of Ref\cite{Kim},
but $\Delta$ and $U$ have been slightly changed from $\Delta=3$~eV and $U=8$~eV
in order to reproduce the observed value of $J$.
$J$ is estimated to be 0.15~eV for these parameters,
and is very close to the observed $J\sim0.16$~eV\cite{Fujisawa}.
The obtained spectra are broadened with a Gaussian function to compare with the experimental spectra.
These are indeed indistinguishable from those in Ref\cite{Kim}.
Contributions from each of the $d$ and $p$ states are shown separately in Fig \ref{calc}.
To compare them with the experimental spectra of Sr$_2$CuO$_3$ of Ref\cite{Fujisawa},
the photoemission cross section effects are considered as shown in Fig.~\ref{calc}(b)
as the total spectra, where the contributions form the $d$-states
are 0.2 times smaller since the He I or Ne I ($h\nu=16.8$~eV)
radiation was used for the measurements.
There are two branches for $0 \le kb/\pi <1/2$,
and only one branch for $1/2\le kb/\pi<1$, agreeing well with
the results of Sr$_2$CuO$_3$ of Ref\cite{Kim,Fujisawa}.
Comparison of the spectra calculated by the $d$-$p$ chain model with
those calculated by the $t$-$J$ model
suggests that two branches in $0 \le kb/\pi <1/2$ are interpreted as a holon and a spinon band,
and one branch in $1/2\le kb/\pi<1$ is interpreted as a holon band\cite{Maekawa}.
The strong peaks, located above 1.0~eV for $k/\pi=0$ and 1/4,
correspond to the $p$-$d$ bonding band.
These peaks could not be clearly observed in the experimental spectra,
since these features overlap with the strong ligand $p$-derived band,
which exists in this region of the experimental spectra. 

Next, we simulate the ARPES spectra of [Ni(chxn)$_2$Br]Br$_2$.
We assume that the difference between [Ni(chxn)$_2$Br]Br$_2$ and Sr$_2$CuO$_3$
is only that in $\Delta$, for simplicity. 
Qualitatively, $t$ and $U$ should be changed for different cation and anion atoms,
but their changes are considered to be not so significant compared with that of $\Delta$\cite{Laan,Zaanen}.
To simulate the ARPES spectra of [Ni(chxn)$_2$Br]Br$_2$, we have estimated
$\Delta$ of this compound in the following steps.
First, we consider the difference in $\Delta$ due to different cation atoms.
In transition-metal oxides, $\Delta$ is increased by about 2.0~eV on going
from CuO to NiO\cite{Bouquet1,CuO}.
The difference in $\Delta$ between CuS and NiS is also estimated to be about 2~eV\cite{Bouquet1},
and therefore we consider that the difference in $\Delta$ from
the difference in the cation atoms is +2~eV.
Next, the valencies of the cations are different between
Sr$_2$CuO$_3$ (Cu$^{2+}$) and [Ni(chxn)$_2$Br]Br$_2$ (Ni$^{3+}$).
In general, it is considered that $\Delta$ of 3$d$ transition-metal oxides is
lowered about 2.5~eV with increasing cation valency \cite{Bouquet1,Bouquet2},
and here we consider that the difference is -2.5~eV.
Finally, the difference in $\Delta$ due to the difference in the ligand atoms are considered.
For the compounds with the same cation, a decrease in ligand electronegativity
will raise the ligand $p$ bands with respect to the metal $d$ orbitals,
resulting in a decrease in $\Delta$.
From the core-level line shape analysis of CuCl$_2$, CuO and La$_2$CuO$_4$,
$\Delta$ is decreased by about 0.8-0.9~eV in going from CuO and La$_2$CuO$_4$
to CuCl$_2$\cite{Okada}.
There is no core-level line shape analysis for CuBr$_2$,
but its core-level spectrum is very similar to that of CuCl$_2$\cite{Laan}.
Hence we assume that the difference in $\Delta$ due to
the different ligand atoms (O and Br) is -1~eV.
Totally, $\Delta$ should be lowered by about 1.5~eV, and therefore we have taken
$\Delta=1$ for [Ni(chxn)$_2$Br]Br$_2$.
The large exchange of $J=0.3$~eV is obtained for the smaller $\Delta$ with same $t$ and $U$,
which agrees well with the larger experimental $J=0.31$~eV of [Ni(chxn)$_2$Br]Br$_2$.
Figure~\ref{calc}(c) shows the calculated spectral function for $t$=1, $\Delta=1$, and $U=7$.
Figure~\ref{calc}(d) shows the total spectra where the contributions
from the $p$ states are 0.2 times smaller
since He II ($h\nu =40.8$~eV) radiation was used for the measurements.
In contrast to Fig.~\ref{calc}(b), only one branch is observed in $0\le kb/\pi \le1/2$,
and it disappears in $1/2<kb/\pi\le1$.
Consequently, the characteristic features of the experimental spectra are
well explained by these parameters, which are consistent with exchange energy $J$.
In these calculations, the hybridization between Ni~3$d$ and N~2$p$ is not considered.
The N atoms of (chxn) molecules are very close to Ni atoms, and,
in addition to the strong crystal field which splits the Ni 3$d$ level,
the hybridization between the Ni~3$d$ and N~2$p$ might influence the electronic structure\cite{Okada2}.
Especially, the spectrum at $kb=0$ might be influenced by this hybridization.
In the calculated spectral functions, there is a strong peak at $\sim1.2$~eV
in the spectrum of $kb/\pi=0$.
This peak is considered to be a $d$-$p$ bonding band,
and may be influenced by this hybridization.

Accordingly, we have successfully described the ARPES spectra of [Ni(chxn)$_2$Br]Br$_2$ and Sr$_2$CuO$_3$
by the same $d$-$p$ chain model Hamiltonian with different charge-transfer energy $\Delta$.
Here, we relate the present results and the strong NLO effects in these compounds.
The NLO effects in 1D SC system have been studied by some theoretical works,
and $\chi^{(3)}$ spectra have been calculated\cite{Saxena,Nasu,Mizuno}.
However, such theoretical approaches require the large cluster size,
and it is difficult to calculate the spectra within the framework of the present model.
Therefore, we qualitatively consider the relationship between
the present results and the NLO effects.
The optical nonlinearity of [Ni(chxn)$_2$Br]Br$_2$ is much stronger than that of Sr$_2$CuO$_3$,
and this difference is ascribed to the difference in the dipole moment ($\langle 0 |x|1 \rangle$)
between the ground state ($|0 \rangle$) and the one-photon-allowed excited state ($|1 \rangle$)
and the dipole moment ($\langle 1 |x|2 \rangle$) between the one-photon-allowed excited state ($|1 \rangle$) and
the one-photon forbidden excited state ($|2 \rangle$)\cite{Kishida}.
The $\chi^{(3)}$ is proportional to the square of $\langle 0 |x|1 \rangle$ and $\langle 1 |x|2 \rangle$,
and therefore the material with larger $\langle 0 |x|1 \rangle$ or $\langle 1 |x|2 \rangle$ give larger NLO effect.
In our analysis, the smaller charge-transfer energy $\Delta$ for the Ni-Br system should make
the $d$- and $p$-levels hybridized well. 
It leads to larger spatial extensions of electron-hole wave functions in the two CT-excited states
$|1 \rangle$ and $|2 \rangle$, and then a larger dipole moment $\langle 1 |x|2 \rangle$. 
Consequently, the smaller $\Delta$ gives stronger NLO effects, and
we predict that [Ni(chxn)$_2$I]I$_2$, which has a lower $\Delta$, should have
stronger NLO effect compared with that of [Ni(chxn)$_2$Br]Br$_2$.

In conclusion, we have studied the electronic structure of [Ni(chxn)$_2$Br]Br$_2$ by ARPES.
The essential features of the spectra are well explained by the $d$-$p$ chain model calculations
with a small charge-transfer energy $\Delta$ compared with that of
the Cu-O based 1D-SC compounds.
This result is consistent with the stronger NLO effect in [Ni(chxn)$_2$Br]Br$_2$,
compared with Sr$_2$CuO$_3$.
We propose that $\Delta$ is an essential parameter for tuning the NLO effects of CT-type
1D-SC electron system.

\acknowledgments
The authors thank Prof. T.~Toyama of Tohoku University and Prof. K.~Nasu and Dr. N.~Tomita of
University of Advanced Study for imformative discussions.

\begin{figure}
\caption{ARPES spectra of [Ni(chxn)$_2$Br]Br$_2$
taken with He II$\alpha$ ($h\nu=40.814$~eV).
An approximate value of $k$ is also shown in each spectrum.}
\label{ARPES}
\end{figure}

\begin{figure}
\caption{Grey-scale plot of the negative second derivetive of the ARPES spectra.
The dotted white line is a guide to the eyes}
\label{image}
\end{figure}

\begin{figure}
\caption{Results of finite-size cluster calculation based on the $d$-$p$ chain model.
(a)Partial spectral function for $t$=1, $\Delta$=2.4, and $U$=7
corresponding to Sr$_2$CuO$_3$.
(b)Total spectral function for Ne I.
(c)Partial spectral function for $t$=1, $\Delta$=1, and $U$=7
corresponding to [Ni(chxn)$_2$Br]Br$_2$.
(d)Total spectral function for He II.}
\label{calc}
\end{figure}


\begin{references}
\bibitem{ARPES}The relation $k=0.5123 \sqrt{E} \sin \theta$ indicates that the 105\%
of the Brillouin zone is covered at the Fermi level with the He II$\alpha$ radiation, while
its only 71\% is covered with He I$\alpha$ radiation.
\end{references}
\end{document}